\documentclass[]{spie}
\usepackage[dvips]{graphicx}
\def\ls{{_<\atop^{\sim}}}
\def\gs{{_>\atop^{\sim}}}
\def\cgs{{ erg cm$^{-2}$ s$^{-1}$}}
\title{A High Resolution Intergalactic Explorer\\ for the Soft
X-ray/FUV}
\author{Martin Elvis\supit{a}, Fabrizio
Fiore\supit{b} (for the CWE Team) \skiplinehalf \supit{a}
Harvard-Smithsonian Center for Astrophysics, 60 Garden St.,
Cambridge MA01238, USA\\ \supit{b} Osservatorio Astronomico di
Roma, Monteporzio, Via di Frascati 33, Rome I-00040, Italy}
\authorinfo{email: elvis@cfa.harvard.edu,
fiore@quasar.mporzio.astro.it} 
\begin{document}
\maketitle

\begin{abstract}
We present a mission concept for high resolution X-ray
spectroscopy with a resolving power, $R \sim$6000,
(c.f. $R\ls$1000 for {\em Chandra}, XMM-Newton).
This resolution is physics-driven, since it allows the thermal
widths of coronal X-ray lines to be measured, and
astrophysics-driven, since 50~km~s$^{-1}$ resolves internal
galaxy motions, and galaxy motions within larger structures.
Such a mission could be small and have a rapid response allowing
us to `X-ray the Universe' using the afterglows of Gamma-ray
Bursts (GRBs) as strong background sources of X-rays, and so
illuminate the `Cosmic Web'. The Cosmic Web is predicted to
contain most of the normal matter (baryons) in the nearby
Universe.

\end{abstract}
\section{INTRODUCTION}
\label{sect:intro}
 
We present a mission concept for high resolution X-ray
spectroscopy at E$<$1~keV with a resolving power, $R \sim$6000,
(c.f. $R\ls$1000 for {\em Chandra}, XMM-Newton).  This resolution
is physics-driven, since it allows the thermal widths of coronal
X-ray lines to be measured, and astrophysics-driven, since
50~km~s$^{-1}$ resolves internal galaxy motions, and galaxy
motions within larger structures. We can then do galaxy dynamics
in X-rays.

As {\em Chandra} and XMM have made clear the region of the
spectrum below 1 keV is where most of the X-ray atomic
transitions lie, and is comparable with the optical and
ultraviolet bands in this richness of features.  However, this
band lags behind optical-UV spectroscopy. High resolution
spectroscopy of this soft X-ray band is the only part of
`discovery space' for which present or future X-ray missions are
not pushing for order of magnitude improvements.  Inoue (2001)
showed how future missions compare to present and past ones in
term of throughput, energy band, and angular resolution. We have
now {\em Chandra} that beats ground-based optical telescopes in
terms of angular resolution (though not HST), XMM-Newton and then
Con-X and XEUS, will have large throughput. Con-X, and other
smaller missions, will image above 10 keV for the first time.
Calorimeters will give high resolution in the iron K band (the
ASTRO-E calorimeter has a resolution of 6~eV, the Con-X
calorimeter could reach 2-3~eV). What is really missing is high
resolution (E/$\Delta E>$5000) at low energy.

Because low energy X-rays only require short focal lengths, such
a mission could be compact and rapidly repointed.  This would
allow us to `X-ray the Universe' using the afterglows of
Gamma-ray Bursts (GRBs) as strong background sources of X-rays,
and so illuminate the `Cosmic Web'. The Cosmic Web is predicted
to contain most of the normal matter (baryons) in the nearby
Universe. A recent flurry of papers has shown that this warm-hot
intergalactic medium (WHIM) does exist, both locally (Nicastro et
al. 2002a b, Sembach et al. 2002) and at moderate redshift
(Zappacosta et al. 2002). The importance of this topic has led to
the approval of SPIDR, a new MIDEX mission. SPIDR will map the
Cosmic Web of warm intergalactic gas in the Far-UV OVI emission
line at 1052\AA, and so will provide strong morphological tests
of cosmic structure formation.

We see a compelling need to go to the next level of the physics
of this major, but elusive, component of our environment: how is
the plasma moving? How is it ionized? Which type of supernova
enriched it with heavy elements?  What is the history of the
formation of the Cosmic Web, and how does this tie in with the
destruction of the `Lyman-alpha forest' of cooler material?  High
resolution soft X-ray and far-UV spectra are the {\em only} means
of studying the detailed physics of the warm gas of the Cosmic
Web.

Gamma-ray Bursts are excellent, though fleeting, background
beacons for these investigations: GRBs explode in galaxies
spanning just the right redshift range, 1$<$z$<$2.  Despite their
location in the distant universe, the X-ray afterglows of GRBs
shine as brightly as the brightest sources in our own local Milky
Way for a short time.  By rapidly slewing a high resolution
spectrometer into position to record the soft X-ray and FUV
spectrum of a GRB afterglow, we can gather 100 times more
photons than missions that point at the only steady
distant sources, high redshift quasars. This will let us study
many absorption lines and resolve their profiles, so telling us
about the physics of the gas in the Cosmic Web.

The GRB soft X-ray/FUV spectra will carry the signatures of {\em
all} material along the line of sight to the GRB, separated out
cleanly by redshift. This will include matter in the GRB host
galaxies, telling us the composition of galaxies during the `Age
of Star Formation' (1$<z<$2, Madau et al. 1996) and so testing
theories of the star formation history of the universe.  The
intimate environs of a gamma-ray burst will also imprint their
signature on the spectra, testing models for these most powerful
explosions in the Universe.

This mission is the complement of the MAP and Planck missions:
they put constraints on one end of the process of cosmic
structure formation; this mission on the other.

\section{The Mission} 

This mission concept is the result of several discussions with a
large group of scientists during two meetings at Johns Hopkins
University\/
\footnote{Contributions by: Andy Szentgyorgyi, Fabrizio
Nicastro, Rob Cameron (SAO), Giovanni Pareschi, Oberto Citterio
(OABrera), W. Cash (U.Colorado), L. Stella (OAR), C. Norman,
J. Rhoads, J. Krolik, S. Savaglio, K. Weaver, T. Yaqoob,
T. Heckman, K. Sembach, D. Bowen, J. Kruk, M.B. Kaiser,
S. McCandless (JHU), A. Fruchter (STScI), S. Mathur (OSU),
J. Greiner, G. Hasinger (MPE). Many thanks to L. van Speybroeck
(SAO), L. Angelini (GSFC), and P. Miotto (Draper Lab.)}  .
The heart of the mission is a high resolution soft X-ray
spectrometer, with 6~times the spectral resolution of {\em
Chandra}. A slitless FUV spectrograph will add the OVI line to
the X-ray OVII and OVII lines, allowing us to distinguish
photoionization from collisional ionization.  A GRB detector will
trigger the spacecraft to slew the X-ray spectrometer onto the
GRB afterglow within a few minutes. (10 minutes requirement, 1
minute goal.)

The soft X-ray spectrometer will consist of: an {\bf X-ray
mirror} (with a 5~arcsec HPD, image size), optimized for low
energy performance ($<$1~keV), feeding photons to {\bf
diffraction gratings} deployed in an out-of-plane reflection
configuration. The gratings disperse the photons onto an {\bf
array of CCD detectors}, which are also optimized for low
energies. The spectral resolution achieved will be $R$=6000
(50~km~s$^{-1}$) over the energy range 0.1--1~keV, with a
collecting area of 1000~cm$^{-2}$ - 2000~cm$^{-2}$. Replicated
X-ray optics developed for Con-X will be the basis of this
design. The short focal length (2.5-3~meters) of the X-ray
telescope mimimizes the moment-of-inertia of the satellite,
allowing faster slews.

To achieve the rapid response necessary to achieve these mission
goals, the satellite will include a compact 2-stage GRB detection
and localization system based on the highly successful Beppo-SAX
design: a {\bf CsI dodecahedron} to localize GRB to
$\sim$1~degree, and a small {\bf X-ray coded aperture telescope}
with a 10-20~degree field of view that can localize bursts to an
arcminute. This is sufficient to put the GRB afterglow into the
(slitless) spectrometer apertures. The X-ray CCDs of the coded
aperture telescope (e.g. an XMM EPIC-pn wafer) will record good
quality X-ray spectra of the afterglows, a bonus for the mission
science.

\section{SCIENCE DRIVERS}

\subsection{The `Missing Baryons' Problem}

It is well known that at high redshifts (z$>$2) most of the
baryons in the Universe lie in `Lyman-alpha forest clouds', at
temperatures of $10^4$ K or less, and with densities only
slightly above the average density of the Universe (overdensities
of only 1-10).  But, in the second half of cosmic time, (i.e. for
the 7~Gyr at z$<$1) the number of baryons in Lyman-alpha clouds
decreases rapidly, while the number of baryons in galaxies and
clusters of galaxies does not increase by the same factor.
Therefore one of the major problems of the modern cosmology is:
{\em ``Where do most baryons go at low redshift?''}

The leading solution to the `missing baryons' problem is the
concept of a `Warm-Hot Intergalactic Medium' (WHIM).  This
concept has been developed in the last few years by Cen \&
Ostriker (1999), and by other groups, (see Dav\'e et al.
2000 and references therein): hydrodynamical simulations of the
evolution of structures in the Universe showed, surprisingly,
that 30-40\% of the baryons at z$<$1 are in a warm ($10^5-10^7$
K) phase at overdensities between 10 and 200.  Most (70\%) of
these baryons should be at only a weak overdensity $<$60, and so
are not virialized and are unbound.
The resulting X-ray `forest' of absorption lines will show us the
structure of the Universe developing over the last 7~Gyr
(Hellsten et al. 1998). Since this structure depends on what
physics causes the primordial fluctuations seen in the Cosmic
Microwave Background (CMB) to grow into the galaxies and clusters
of galaxies we see today, the X-ray forest will constrain that
fundamental physics.

A {\em Chandra} spectrum was recently used to discover X-ray
absorption at zero redshift due to the WHIM (Nicastro et al.,
2002a), following years of frustrating upper limits (Aldcroft et
al. 1994).  The WHIM thus does exist, at least locally and along
on a single line of sight, and with the predicted overdensity.
The bare detection of collisional OVI, OVII and OVIII lines is a
prime scientific result, strongly corroborating the results of a
decade of forefront cosmological modeling.  Most likely the IGM
component found by Nicastro et al. is a cut through the `local
filament' associated with the Local Group of glaxies (which is
dominated by the Milky Way and Andromeda spirals.)

Examined in this light FUSE detections of zero redshift
absorption toward many AGN implies that we are embedded in a WHIM
filament of the Cosmic Web. Nicastro et al. (2002b, see also
Sembach et al. 2002) demonstrate that these OVI absorbers are at
rest in the frame of our Local Group of galaxies, but not in any
Milky Way frame of reference. This clinches the extragalactic
location of this highly ionized gas. Moreover, the mass of gas
implied is enough to bind the Local Group gravitationally, which
begins to put interesting limits on the amount of `dark matter'
in our vicinity.

\begin{figure}
\includegraphics[height=7cm]{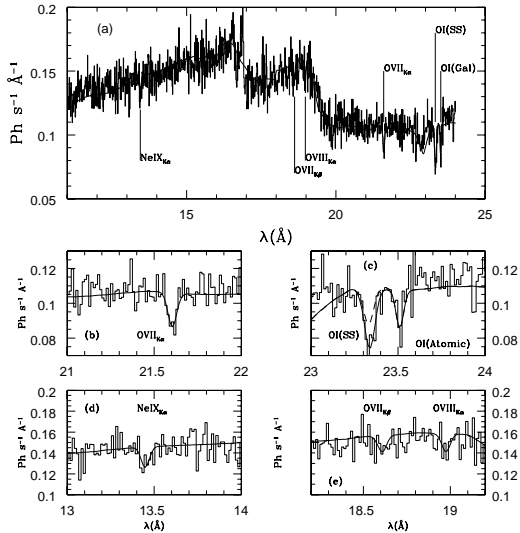}
\caption{Chandra discovery of IGM absorption lines (Nicastro et
al. 2002a). Residuals after subtraction of the continuum of the
spectra, in velocity space, of the OVII$_{K\alpha}$, NeIX, OVIII,
OVII$K\beta$, and OI$_{1s-2p}$ resonant lines, from the LETGS
spectrum.}
\label{nicastro}
\end{figure}

To understand the structure of the warm IGM making up the Cosmic
Web will require many more directions to be probed to distances
well beyond our parochial group of galaxies.  Observing out to
z=2 along many lines of sight will let us see the scales on which
the Web has formed and will show us how the Web has grown and
been filled with X-ray hot gas, gradually replacing the cooler
`Lyman-alpha forest' gas. As we show later, quasars are a poor
choice of background source; GRBs are far better.

The breakthrough science of the Cosmic Web will come from {\em
measuring the widths of the oxygen lines}. This will allow a
series of new tests of cosmological models For example,
hydrodynamical simulations suggest that the IGM matter is heated
to $10^5-10^7$ K by shocks during the collapse of density
perturbations. If this is the case one would expect that the
turbulent velocities in the shocked gas of the same order of
magnitude as its sound speed, i.e. of the order of 100-200 km/s.
This is a several times the thermal velocity of the heavier
oxygen atoms (at $4\times10^6$ K, the temperature at which the
baryon density peaks in hydrodynamical simulations, Dav\'e et
al. 2000). Detecting line widths of 100-200~km~s$^{-1}$ would
point toward shocks as main source of large scale heating
of the gas. This would contrast with competing sources such as
supernova heating, and so determine the heating mechanism.  GRB
soft X-ray/FUV afterglow spectroscopy will allow these tests to
be carried out.

\begin{figure}
\includegraphics[height=7cm]{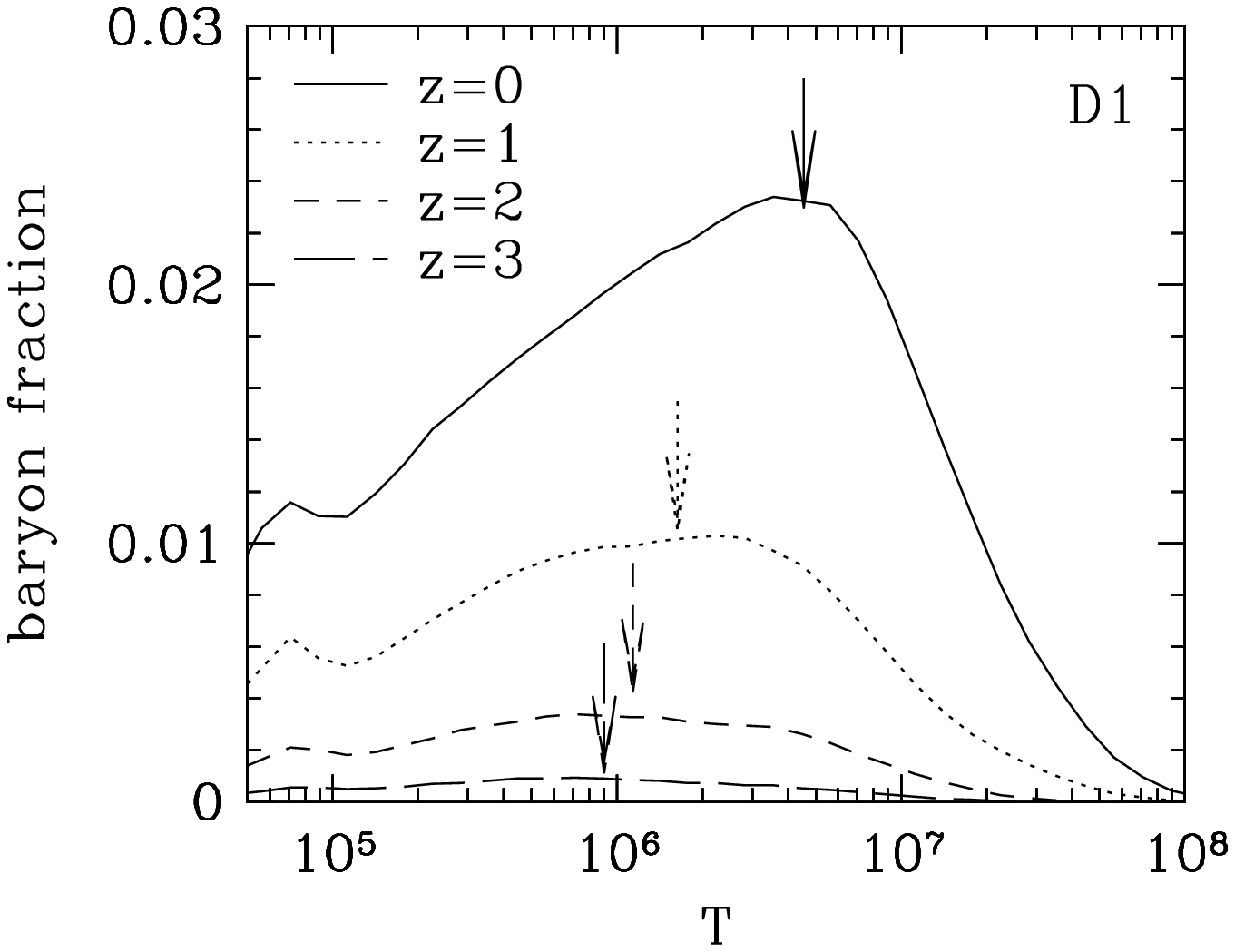}
\includegraphics[height=7cm]{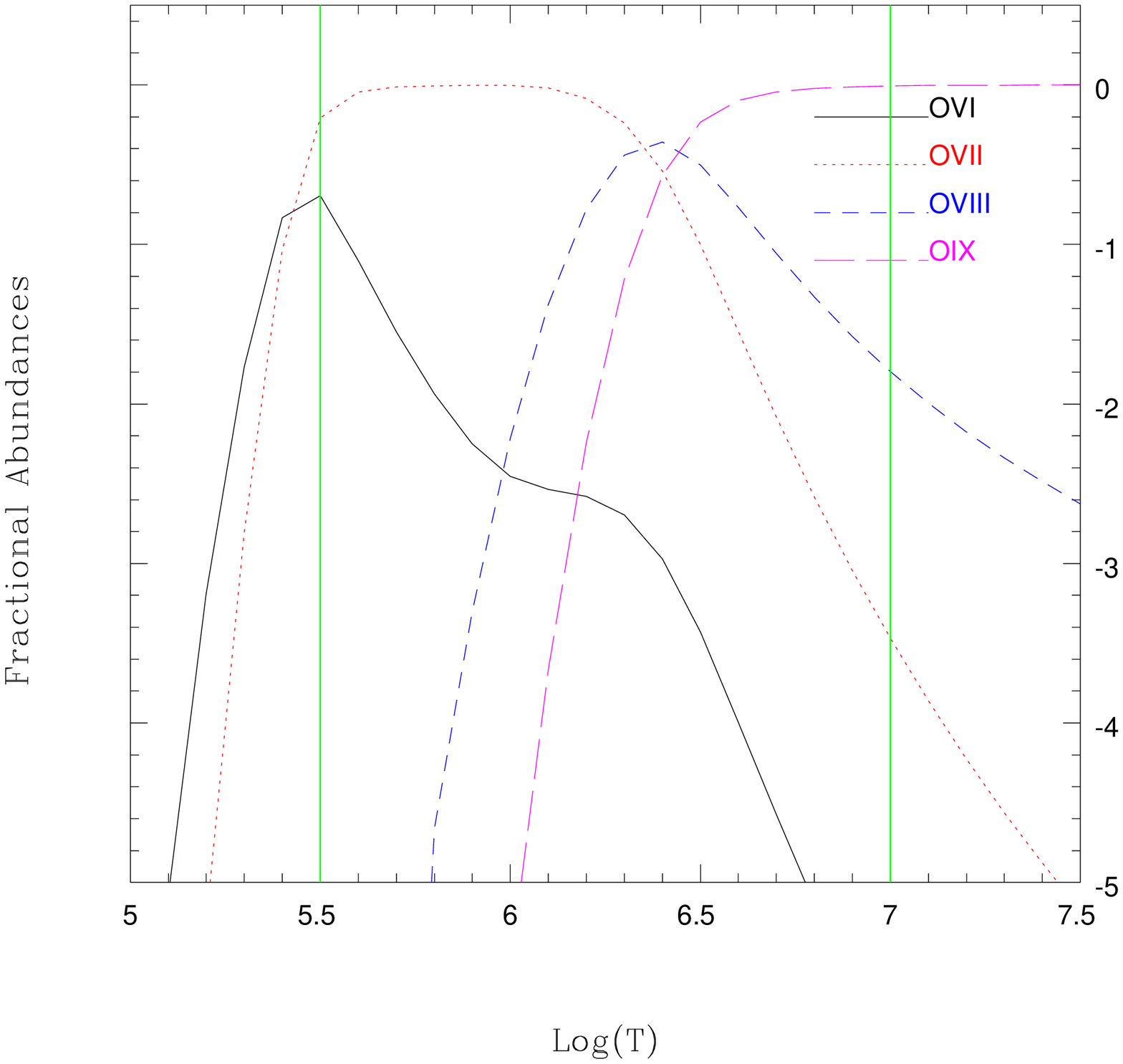}
\caption{{\em left:} the baryon density as a function of the
temperature (Dav\'e et al. 2000). {\em right}: the oxygen ion
fraction as a function of the temperature, assuming collisional
equilibrium. Adapted from Dav\`e et al. (2001).}
\label{compare}
\end{figure}


\subsection{Galaxies in the Age of Star Formation}

Prompt observations of GRB afterglows offer a new and distinctive
path for the study of the matter in the immediate surroundings of
the GRB (r$\sim$100~pc) and in the GRB host galaxy (Fiore et
al. 2000). This is the second main science driver for this
mission. Around $z=2$ is the age of star formation (Madau et
al. 1998), and many GRBs are found at such redshifts.

As a result GRBs can be a powerful tool to determine the history
of the metal enrichment in galaxies in the Universe (Fiore 2000,
2001, Savaglio, Fall \& Fiore 2002), which can then be compared
with enrichment predictions of theoretical models (e.g. Cen \&
Ostriker 1999), to pin down their several, now unconstrained,
assumptions.

GRB host galaxies appear typical of normal star-forming field
galaxies at the same large redshifts (Bloom et al. 2001,
Djorgovski et al. 2001). Moreover GRBs occur well within the main
body of their host galaxies, not in the outer haloes. Best of
all, since GRB host galaxies are $\gamma$-ray and X-ray selected
they will be virtually unbiased against dusty environments, a
serious limitation of all present studies of high~z galaxies.

Optical spectra demonstrate clearly that GRB observations can be
used to probe the ISM of the GRB host galaxy. Using a GRB
afterglow spectrum, Castro et al. (2001) discovered two
absorption systems in the GRB host galaxy separated by just
$\sim150$~km~s$^{-1}$. But optical observations are unable to
investigate the abundant hot material expected. Neither can
normal ultraviolet, or infrared, telescopes detect this
material. Only the far-UV and soft X-ray bands contain the
necessary spectral signatures.


\section{Spectroscopy Goals}

The soft X-ray band will measure OVII and OVIII, while
the FUV band will measure OVI. The ratios of these lines will
determine the temperature of the WHIM for each line of sight
(figure 2). Crucially, the two line ratios come from 3 ions of
the same atom. If the ratios disagree on the temperature, then we
will know that we are not dealing with a simple collisional
plasma (Nicastro et al. 2002a).

The soft X-ray spectra will also contain NeIX, while the FUV
contains iron, magnesium, silicon, silicium, carbon and zinc and
hydrogen Lyman-$\alpha$ lines. With the oxygen ratios
distinguishing unambiguously between photoionization and
collisional ionization, these other lines will give us abundances
and enrichment histories. This will tell us the history of
supernovae, and of the recycling of matter from galaxies and
quasars. Some elements condense onto dust grains more easily than
others (Pettini et al. 1997, 1999), so the dust content of the
universe will also be measured. Dust is a catalyst for further
star formation, and so the WHIM dust content should link with the
star formation history of the Universe. Dust dims light from more
distant objects, and allowing for this (probably small) effect
might change the cosmological parameters derived from SN1a light
curves.

\section{Why GRBs are the Best Path}

The spectroscopic goals above are challenging. They require a
large number of photons in each spectrum.  We can use the example
of the Chandra LETGS/HRC PKS~2155-305 spectrum of Nicastro et
al. (2002a) to determine how many photons are needed to obtain
sufficient signal-to-noise.

\begin{figure}
\label{ctsigma}
\includegraphics[height=7cm]{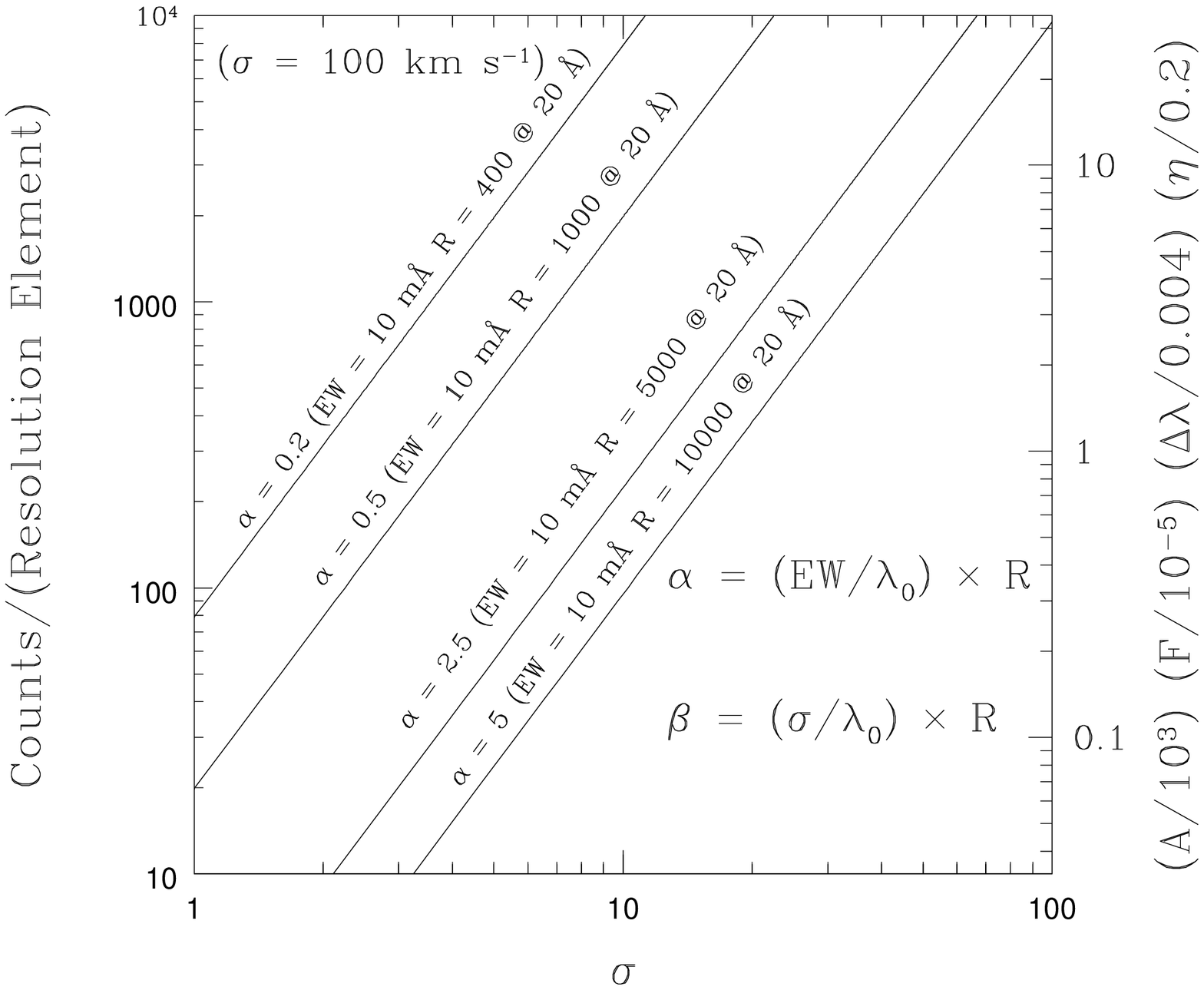}
\includegraphics[height=7cm]{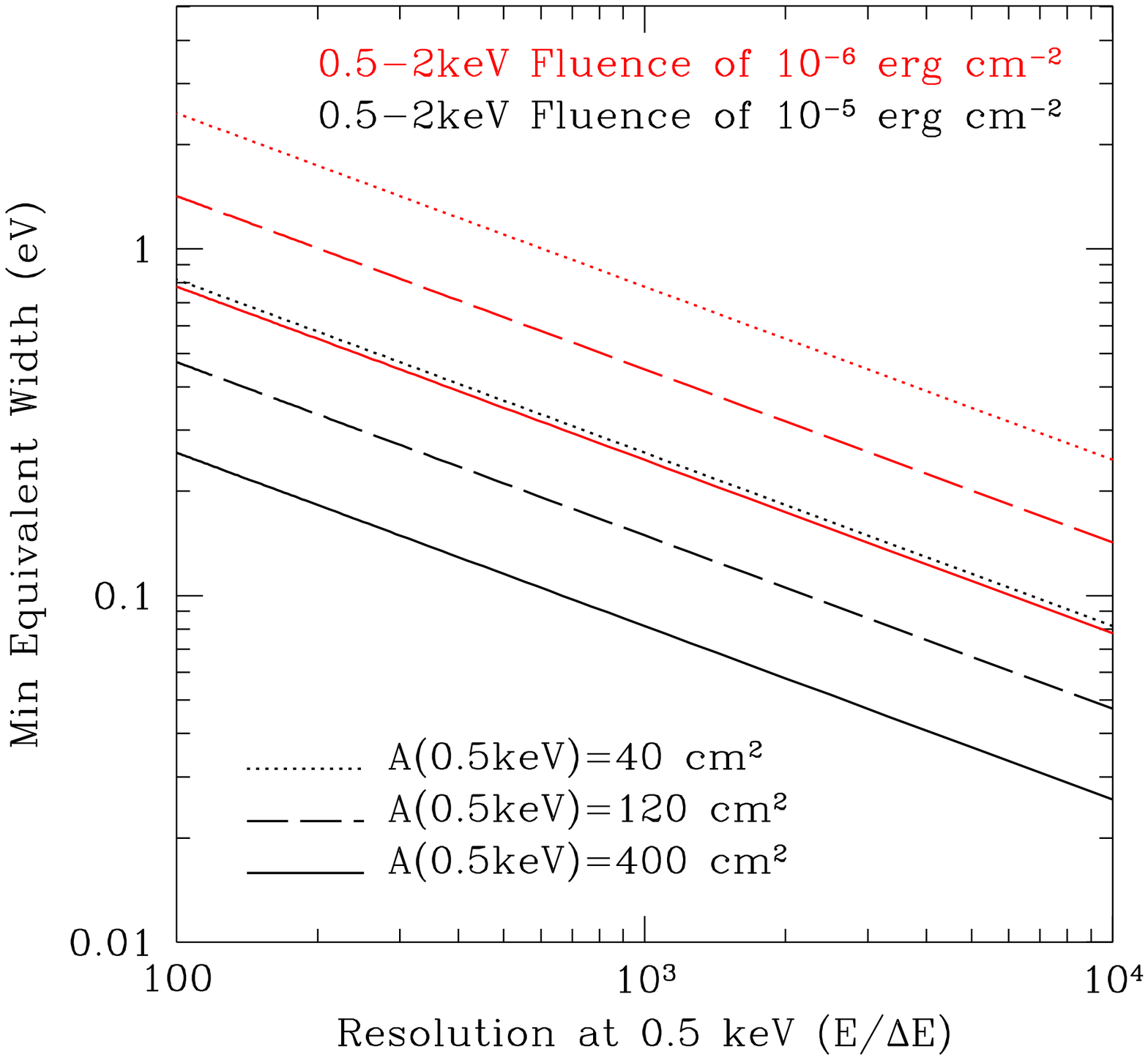}
\caption{{\em Left:} Number of counts per resolution element
(left vertical axis) vs. detection significance, $\sigma$. The
right vertical axis shows the Area, $A$, fluence, $F$,
resolution, $\Delta\lambda$, efficiency, $\eta$, product needed
to obtain the corresponding number of counts per resolution
element. {\em Right:} Minimum detectable absorption line
equivalent width vs. resolving power, $R$.}
\end{figure}

The Chandra LETGS/HRC PKS~2155-305 spectrum has between $\sim$600
and $\sim$1200 counts per resolution element. The most intense
lines detected in PKS~2155-304 have an equivalent width,
EW=10.4~mA. The resolution element with the line centroid
contains $N(line)$= 511~counts, and the continuum around the line
has instead $N(cont)$= 692~counts/res.element. The difference
$D(centroid)$= [$N(cont)$ - $N(line)$] = 181~counts/res.element.
To detect such a line at $M\sigma$ we need at least $N(line)$
such that $D(centroid) >= N(line)[1 + M/\surd{N(line)}]$.  $D$
depends both on the EW of the line and on the resolution.  For an
instrument with a resolving power $R = 6000$, $N(cont)$ is
reduces so $D$ is bigger for a given EW, and a detection at a
given $M\sigma$ is easier to obtain (figure~\ref{ctsigma}).

To resolve a line, the above computation has to be scaled for the
line profile in the contiguous resolution elements, with 3-5
elements being minimal. For a gaussian with a FWHM of
100~km~s$^{-1}$, at 21.6\AA\ and an instrument with a resolution
of 6000 at the same wavelength, the line would be resolved into
roughly 3-5 resolution elements. For a gaussian the 2 resolution
elements either side of the one containing the centroid would
each contain roughly 15\% of the total number of counts.  This is
about 20\% of the number of counts contained in the
centroid-resolution element. So, $D(wings) \sim 0.2 D(centroid)$.

BeppoSAX results showed that at least 7 out of 11 SAX GRB with
fluence (i.e. integrated flux) $>$1$\times$10${-5}$ have bright
X-ray and optical afterglows (Fiore 2001). Hence these bursts
will provide the photons that will yield the high quality spectra
needed to carry out the science.  Because GRB X-ray afterglows
can be so bright, prompt GRB observations (i.e. within one or a
few minutes - a {\em Swift}-like response) provide a huge
advantage compared to bright quasar observations, by providing
large fluence of X-ray photons (Fiore et al. 2000).  For example:
a 40ks (1/2 day) observation of a GRB with the same peak X-ray
flux as GRB990123 or GRB010222 (In't Zand et al. 2001, Jha et
al. 2001, Masetti et al. 2001) that begins 1~minute after the
burst onset provides a fluence (10$^{-5}$~erg~cm$^{-2}$,
0.5-2~keV band) equivalent to a one million second (about
2~weeks) long observation of a bright ($F_X\sim10^{-11}$ \cgs, or
0.5 mCrab) z=1 AGN. At z$>$0.2 there are only a dozen AGNs in the
sky as bright as this, while there are several such GRB each
year.

\begin{figure}
\label{lnls}
\includegraphics[height=7cm]{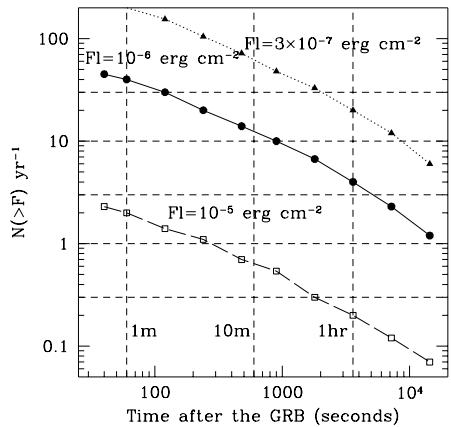}
\caption{Number of GRBs (in 40\% of the sky) per year with X-ray
fluence (integrated number of photons) greater than 10$^{-6}$ and
10$^{-5}$~erg~cm$^{-2}$ (0.5-2~keV), as a function of time after
the burst. The figure was produced using the recipes of Fiore et
al. (2000), and assuming an afterglow decay index of -1.3 and GRB
rest frame absorption of $10^{21}$ cm$^{-2}$.  (A fluence of
$10^{-5}$ erg~cm$^{-2}$ is equivalent to an observation of a
$f_X\sim10^{-11}$ \cgs AGN for 1 million sec.) The importance of
reaching a GRB within a few minutes is clear from the rapid
decline of fluence with time.}
\end{figure}

Figure \ref{lnls} gives the number of GRB per year (at high
galactic latitude) which give a 0.5-2 keV fluence (in a 40~ks
observation) of $10^{-5}$ and $10^{-6}$ erg cm$^{-2}$, as a
function of the delay time between the GRB and the start of the
observations.  The GRB keV logN-logF of figure~\ref{lnls} (using
a slope of -1.3) can give the total X-ray fluence from bright
bursts ($fluence(X)\gs 10^{-6}$ erg cm$^{-2}$) per year.  If 40\%
of the high Galactic latitude GRB are included (since about half
will be occulted, or caught too close before occultation, by the
Earth for a LEO satellite) with a delay of one minute, this gives
a total fluence of about $\sim4\times10^{-5}$ erg cm$^{-2}$,
equivalent to a 4 million seconds observation of a
$F_X\sim10^{-11}$ erg~cm$^{-2}$~s$^{-1}$ AGN.  A HEO satellite
would record double this total fluence. If the delay time rises
to 10 min, then the total fluence would be reduced by two thirds.

In the optical and UV bands observations of GRB afterglows are
probably less efficient with respect to quasars for the study of
the IGM, since quasars have a much bluer spectrum than GRB
optical afterglows. There are about 100 quasars with UV flux
$>10^{-14}$ \cgs Hz$^{-1}$, accessible to FUSE, while to obtain
spectra of good quality (S/N$>$20 per resolution element) of a
GRB afterglow needs an instrument with an effective area about 10
times that of FUSE. However these 100 quasars would need a {\em
much} larger X-ray telescope to give the OVII and OVIII
lines. Only GRBs give both easily.

There will be about 20 GRBs/year with $fluence(X)=1\times
10^{-7}$.  These bursts provide the driving science of the
mission - velocity resolved absorption lines of the WHIM. They
will determine the state of the X-ray forest lines.
In addition to the primary GRBs there will be 80/year that are
faint ($fluence(X)=3\times 10^{-7}$. For these we will
detect strong lines at redshifts inaccessible to SPIDR.
This will give us a robust statistical view of the
formation of the Cosmic Web with time.
About once per year a bright ($fluence(X)=1\times10^{-5}$) GRB
will be caught, allowing resolved velocity structure to be seen
even in faint lines, which will give much tighter constraints on
nucleosynthesis in the Age of Star Formation.


The X-ray coded aperture GRB location instrument will collect a
$\sim$5000 count CCD spectrum for a GRB of medium fluence. This
is sufficient to detect an Fe-K line at the GRB redshift, freeing
us of the need for optical ground-based follow-up (although this
would still be valuable).  Changes in Fe-K strength with time
seem to be strong (Reeves et al. 2002), and can be studied with
the mission.


The combined telescopes make up a powerful instrument package
that will be able to study the IGM on a whole range of scales.
Since we will follow about 120 GRBs per year, and each one
will be detectable for no more than one day, the satellite will
spend a sizeable fraction of the time, 240 days/year, observing 
the IGM on more local objects:

\noindent{\bf z$<$0.2, Filaments between Nearby Clusters:} Long
observations of the 10-20 brightest AGN in the sky with
z$>$0.2. Each of these observations will take about 1~month, (for
50\% efficiency in LEO), and so will fill a 2-year secondary
science mission. Triggers from X-ray all sky monitors may allow
us to catch some of these AGN in outburst, and so with fluxes
5-10 times normal. Most AGN have `warm absorbers' similar to the
IGM, but denser. They form a fast wind emanating from the AGN,
and their large outflow velocities can be confused with
intervening matter. Fortunately during this month of monitoring
most AGN vary significantly, which will cause changes in any
absorbing material close to the AGN, distinguishing them clearly
from the IGM. These variations will determine the density of AGN
winds, a valuable byproduct of the mission.

\noindent{\bf Damped Lyman-$\alpha$ Absorbers (DLAs):} DLAs may
be protogalaxies (Prochaska \& Wolfe 1999).  Soft X-ray/FUV
spectra will determine the abundances and enrichment processes of
DLAs (Bechtold et al. 2001).

\noindent{\bf The Local Group Filament:}
Observations of X-ray sources in the Local Group (the Magellanic
Clouds, M31, M81/M82, M33) and of the Milky Way halo via globular
cluster X-ray binaries, will separate out the warm gas in our
galaxy halo from gas in the IGM filament in which our galaxy
and the Local Group of galaxies lie.


\section{INSTRUMENTATION}

\subsection{Energy/Wavelength Coverage}

All the interesting X-ray spectral features are at E$<$2 keV,
even in the rest frame, and for z$>$1 they all lie at E$<$1~keV
(figure~5). Hence the primary science goals of the
mission require that the X-ray spectra should reach to as low an
X-ray energy as possible ($\sim$0.1 keV), but need not reach to
higher energies than 1-2~keV.

The FUV allows the study of the OVI~1054\AA\ and
Ly$\alpha$1215\AA\ lines. The OUV band (1500-4000\AA) is needed
if we are to study lines of several elements including: carbon,
silicon, iron and zinc.

\begin{figure}
\label{lines}
\includegraphics[height=7cm]{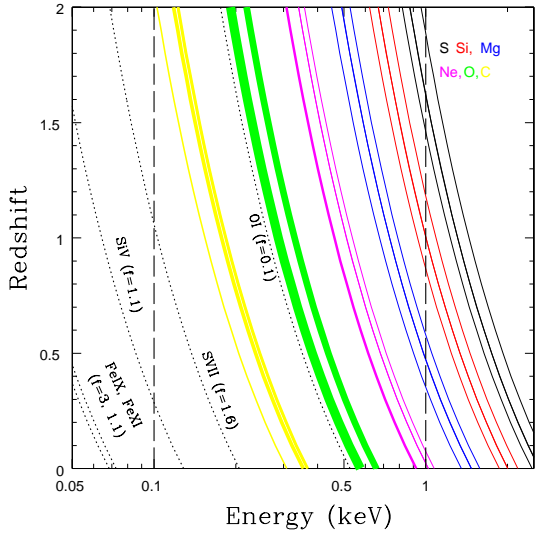}
\caption{Major warm-hot intergalactic medium X-ray absorption
lines vs. redshift, $z$. A nominal observing band of 0.1-1~keV is
shown.}
\end{figure}

\subsection{Spectral Resolution}

The resolution of a spectrometer is normally measured in terms of
{\em resolving power}, which is the wavelength divided by the
FWHM of the instrumental line width:
$R$=$\lambda/\Delta\lambda$. 


A resolving power of $R\sim$400 was just sufficient to detect the
strongest absorption lines from the WHIM (Nicastro et al. 2002a),
but is not enough to pin down cosmological models from their
detailed properties.  A clear example of the detail lost at
$R=1000$ is given in figure~6, which compares in velocity space
the absorption lines detected by FUSE and Chandra LETGS-HRS for
the BL~Lacertae object PKS2155-301 (Nicastro et al. 2002a).  At
the FUSE resolution the OVI system is resolved in at least two
components, one narrow and one broad. Although this complex
system is detected with a very high significance in the soft
X-ray spectrum the 10 times lower resolution does not resolve the
components. A resolution of 6000 ($\Delta v$=50~km~s$^{-1}$) is
needed to resolve the thermal oxygen lines (for T$\gs4\times10^6$
K, see also Elvis 2001).

The FUV allows the study of the OVI~1054\AA\ and
Ly$\alpha$1215\AA\ lines.  Here a resolution of $R$=10,000,
comparable to FUSE, would allow the detection and the
characterization of faint oxygen lines. A resolution of $R$=3000
is barely enough for the detection of strong OVI lines, but would
allow detection of Ly$\alpha$ absorption lines.

\begin{figure}
\label{2155}
\includegraphics[height=7cm]{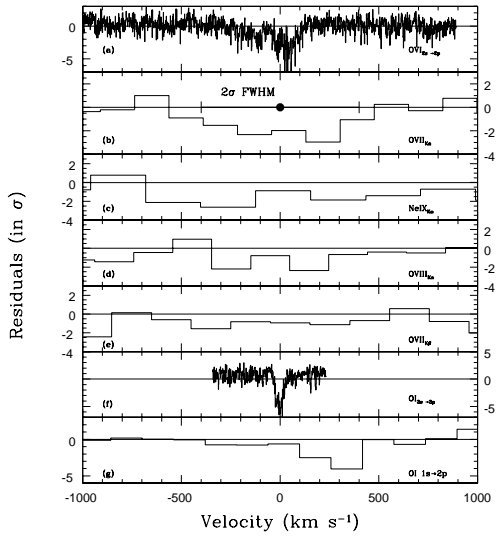}
\caption{Residuals after subtraction of the continuum of the
spectra, in velocity space, of the OVI (FUSE), OVII$_{K\alpha}$,
NeIX, OVIII, OVII$K\beta$ (Chandra LETGS), OI$_{3s-3p}$ (FUSE)
and OI$_{1s-2p}$ (Chandra LETGS) resonant lines.  Adapted from
Nicastro et al. (2002a)}
\end{figure}

\subsection{Rapid GRB Trigger and Localization}

To obtain spectra of GRB afterglows, first one must find them,
quickly. Moreover there are only about 100 bright GRBs at high
Galactic latitude per year, so few can be missed.
Hence the initial GRB trigger detector needs the largest possible
field of view: 2$\pi$-4$\pi$ steradians; yet to use the X-ray/FUV
spectrometers requires a position accurate to at least an
arcminute. These are hard requirements to reconcile in a single
instrument.
By concentrating solely on X-ray bright GRBs we can divide the
problem into two parts: (1) triggering on a GRB detection, and
(2) localizing the GRB accurately enough for the soft X-ray/FUV
spectrometers.  Moreover, since this is a soft X-ray spectroscopy
mission the GRB localization instrument only needs to operate in
the typical 0.5-10~keV band of CCD detectors. The high energy
response of BAT on {\em Swift} is not needed. In this way the GRB
detection and localization system can be far more compact and
light weight than that of {\em Swift}. ({\em Swift} will pursue
{\em all} GRBs.)  Since {\em Swift} can slew 1~radian in a
minute, our much smaller and lower moment-of-inertia satellite
should have no problem slewing 2~radians in a similar time.

\noindent (1) {\em Trigger:} A GRB monitor (GRBM) similar to (but
less sensitive and $\sim10$ times smaller than) BATSE, and
capable of providing positions with accuracies of 1-10
degrees. The CsI GRB alert monitor on Beppo-SAX provides a good
model.  The spacecraft autonomously decides whether each burst
detected satisfies the trigger criteria for a re-pointing of the
spacecraft, and then slews to the rough GRB position.

\noindent (2) {\em Localization:} An X-ray Wide Field
Coded Mask (WFCM) with a CCD detector (e.g. an XMM EPIC-pn wafer)
provides a fine position.  A field of view 2-3 times the size of
the GRBM positional error regions ($10\times10$ or $20\times 20$
degrees) is sufficient to safely cover the error box provided by
the GRBM, which the spacecraft will have placed near the center
of the WFCM field of view.  This relatively small field of view
(c.f. BAT on {\em Swift}) allows the use of pixels small enough
to provide positions accurate to within 1~arcmin.  The WFCM
provides the improved GRB position to the spacecraft to perform a
fine maneuver to put the high resolution spectrometers onto the
target GRB.
The WFCM needs sufficient signal to locate the burst in a few
seconds, else valuable fluence is lost.  The minimum area needed
to obtain such a position in a few seconds is somewhat smaller
than that of an EPIC-pn chip ($\sim36$ cm$^2$).  The computation
of the GRB position will be greatly speeded up since there will
be only a single source producing mask shadows on the detector.

\subsection{X-ray Mirror Area}

Our science goals show that we need to design an X-ray telescope
providing an effective area of at least 1000~cm$^2$ below
1~keV. The mirror needs a relatively sharp PSF
(HPD$\sim$5~arcsec) in order to produce high spectral resolution
using gratings.
There are reasons for optimism that 5~arcsec HPD can be obtained.
Silicon carbide shells built by the Merate (Italy) group
(Citterio, Pareschi and collaborators) have 11~arcsec HEW figure
errors.  These shells are 2mm thick, 60cm diameter, 3.5m focal
length, and have a weight/effective area ratio of 0.06
kg~cm$^{-2}$. Mandrel quality is expected to improve (Pareschi et
al. 2001), and the HEW should roughly scale linearly with the
thickness down to about a HEW of 5 arcsec, where other effects
start to be more important than the mirror deformations.

For a small mission weight is a major constraint. As a rough
estimate the mirror will need a support structure of similar
weight. The detector and its associated electronics will likely
have a similar mass, and the total payload will likely be 50\% of
the total mass to orbit, the spacecraft taking the remaining
fraction. The mirror should then account for about 10\% of the
total to-orbit mass. For a MIDEX the mass to orbit is about
1000~kg), this implies a mirror mass of $<$100~kg.
We have explored the mirror parameter space with the {\tt mirror}
raytrace code kindly provided by Leon Van Speybroeck. This code
was used extensively for {\em Chandra} design and development. We
show here the results of three representative design options:
thin vs. thick shells, short vs. long focal length, and iridium
vs. nickel coating. They give a feel for the effective areas
which the X-ray mirror can achieve within the parameter space.
The effective area results for simulations of mirror designs
assuming a `minimal' mass telescope, and a `maximal' mass
telescope (figure~7). In both designs various
combination of mirror coatings have been used, with Ni being used
on the outer shells and Ir on the inner shells.  Nickel provides
better reflectivity below 0.8~keV, and iridium above 0.8~keV, so
the mixed coating designs use iridium for the inner shells, and
nickel for the outer shells.

%

\begin{figure}
\label{mirror}
\includegraphics[height=7cm]{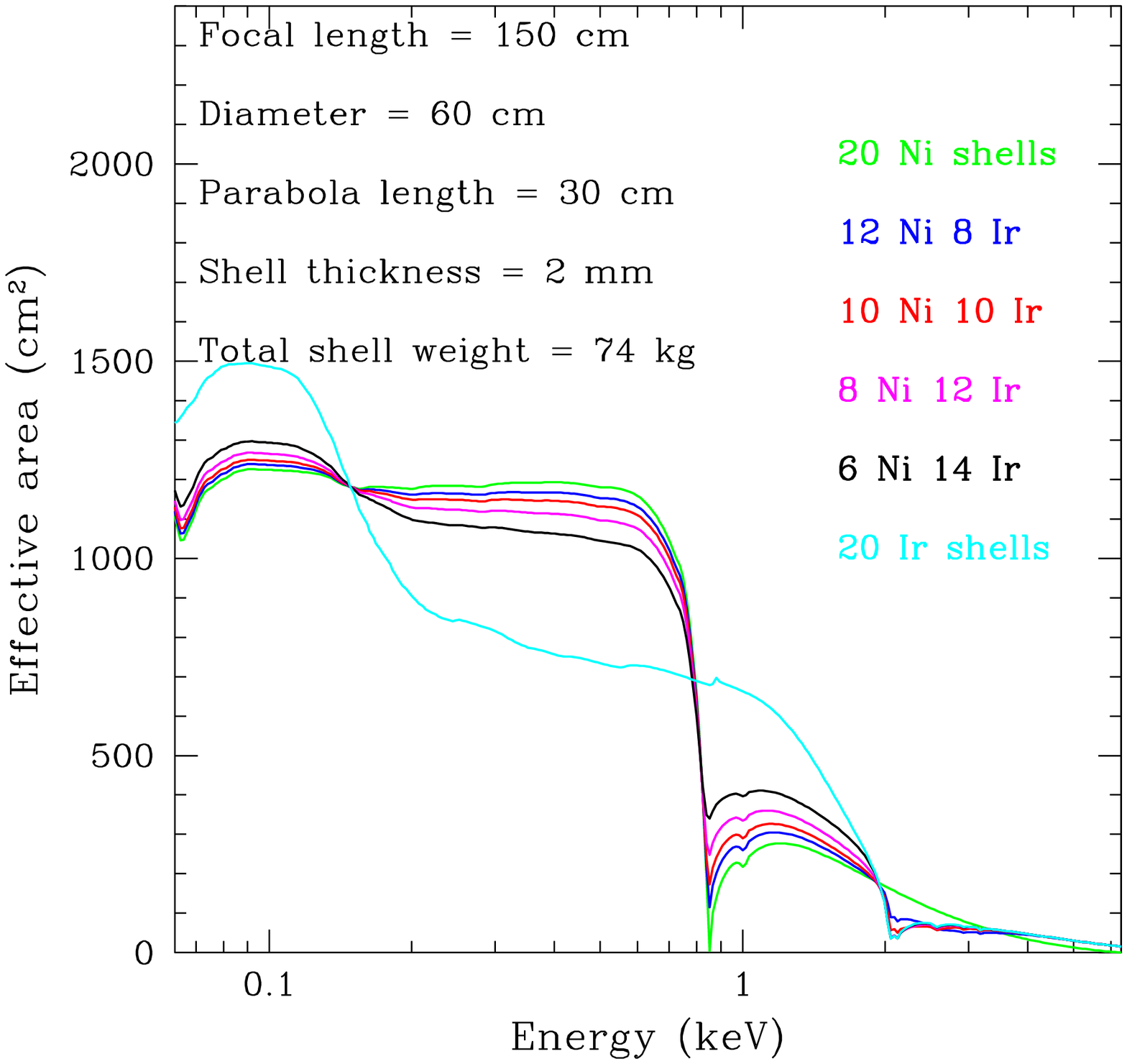}
\includegraphics[height=7cm]{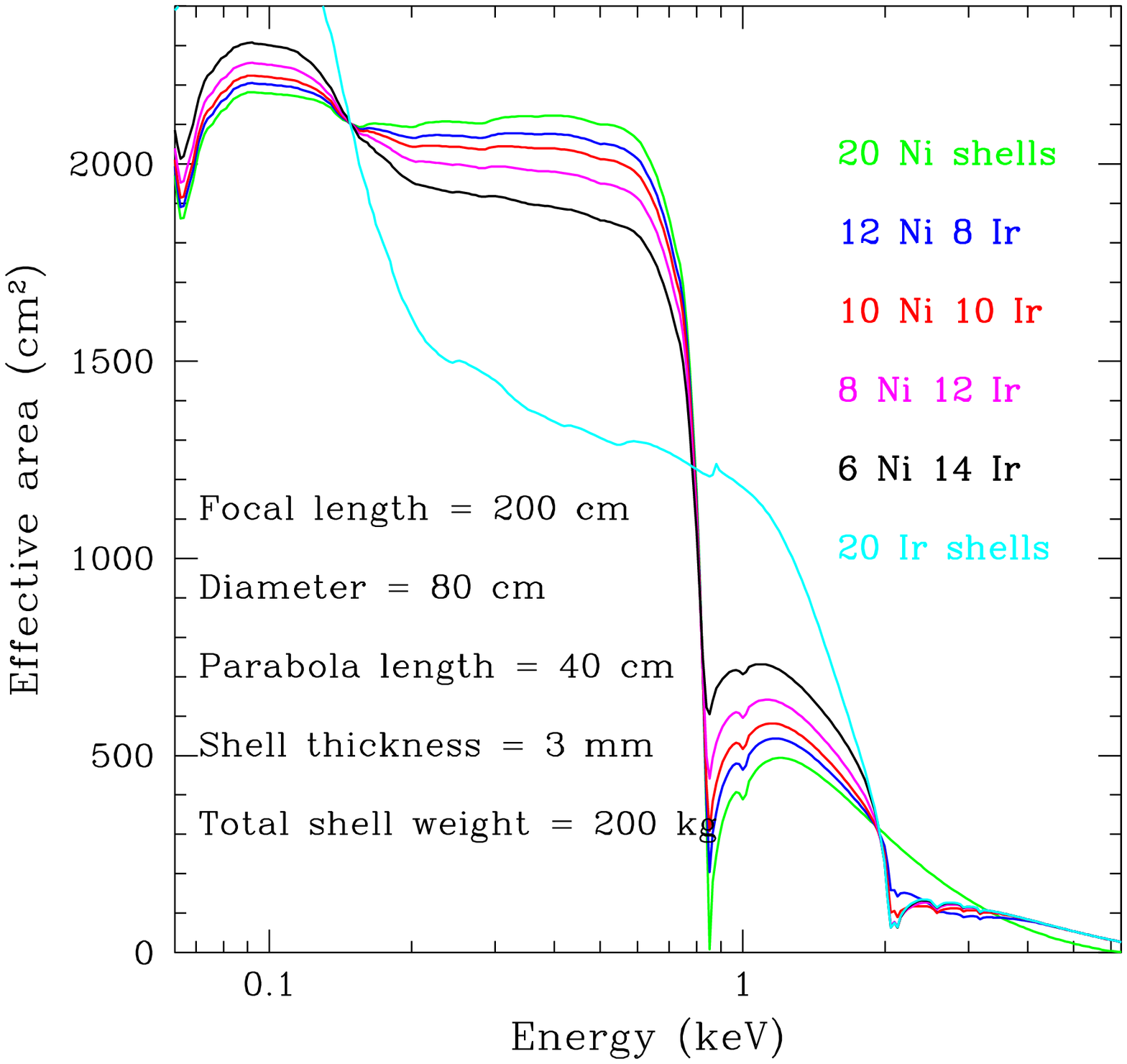}
\vspace{-2cm}
\caption{{\em left:} Raytraced mirror effective areas for a
`minimal' mirror design based on 20 2mm thick shells, 1.5~m focal
length, 60~cm outer mirror diameter, and 30~cm parabola length.
The total mirror weight (including a 100\% support structure
margin) is 96~kg. {\em right:} `maximal' mirror design: 20 3mm
thick shells, 2~m focal length, 80~cm outer mirror diameter, and
40~cm parabola length. The total mirror weight (including a 100\%
support structure margin) is 260~kg.}
\end{figure}

\subsection{X-ray Gratings}

The high spectral resolution required by our science drivers is
technically feasible. If the XMM-Newton RGS reflection grating
facets were used behind the {\em Chandra} mirror the result would
be $R\sim 5000$ (subject to better facet alignment, Elvis 2001).
But is $R=6000$ feasible in a small mission?

X-ray calorimeters have excellent quantum efficiency and are the
natural spectrometer of choice at the 6.4~keV Fe-K
complex. However they are not a good choice at 0.5~keV, since
their resolution is fixed in eV, so $R$ degrades linearly with
energy. The Con-X calorimater goal is to have $\Delta E$=1~eV, or
$R$=500 at 0.5~keV. Gratings are a requirement for this mission.

A route to achieving $R=6000$ with 5~arcsec HPD mirrors is
offered by out-of-plane reflection gratings (ORGs, Cash 1991).
ORGs are being considered to boost the resolution of the Con-X
low energy spectrometers to $R$=1000-2000. Spectral resolution
improves linearly with angular resolution. So with a factor 3
better mirror, and an optimized alignment, $R=6000$ becomes
feasible.  ORGs also have favorable efficiency, peaking at
$\sim$10-15\%. (c.f. 4\% of the XMM RGS, and the 10\% of the {\em
Chandra} LETGS), so requiring a smaller mirror to feed them
photons.

Filters are needed to eliminate stray optical light. To avoid
losses at and above the 0.28~keV Carbon edge, we will use
aluminum filters, which have an edge at 1.5~keV, above our
primary range of interest. Similar filters have been successfully
flown on other missions.

\section{Figure of Merit: Comparison with other Missions}

A figure of merit (FoM) for a system mirror+gratings can be given
by the following formula:
\begin{equation} 
FoM=A_{eff}(cm^2) \times \epsilon_{peak} \times R(0.5 keV).
\end{equation} 
This FoM measures the ability of various systems to detect faint
absorption lines, since the minimum detectable line equivalent
width, $EW_{min}\propto\surd{R(eV)/A_{eff}}$.  Table~\ref{fom}
shows that the mission we propose represents a large gain over
existing or planned missions, even for warm IGM {\em
detection}. Higher resolution spectroscopy not only allows the
line detection, but is essential to the study of the physics and
the dynamics of the absorbing systems. {\em No other mission} can
resolve the X-ray absorption lines, and so take the subject to
the next level of physics and cosmology.

\begin{table}
\caption{Comparison of Absorption Line Detection Capability of
Relevant Missions}
\label{fom}
\begin{tabular}[h]{|lcr|}
\hline
Mission& A$_{eff}$(cm$^2$)$\times$
      R(0.5~keV)$\times$$\epsilon_{peak}$& FoM\\  
\hline
Current Missions&& \\
XMM 1 RGS& 70$\times$ 300$\times$ 0.1& 2100 \\
Chandra LETGS& 100$\times$ 500$\times$ 0.1& 5000 \\
Chandra MEG& 75$\times$ 1200$\times$ 0.1& 9000 \\
\hline
Future Missions&&\\
{\bf `minimal' mission}& 1000$\times$ 6000$\times$ 0.1 &600,000 \\
{\em `maximal' mission}& 2000$\times$ 6000$\times$ 0.15&1,500,000 \\
Swift& 100$\times$ 10$\times$ 1& 1,000 \\
Con-X 1 grating& 1250$\times$ 500$\times$ 0.1& 62,500 \\
Con-X 4 gratings& 5000$\times$ 500$\times$ 0.1& 250,000 \\
\hline
\end{tabular}

Note: includes response time estimate of 4~hours for missions
other than this, and 1~minute for this mission, assuming a
power-law decay index of $-$1.3.
\end{table}

\section{Conclusions}

High resolution ($R\ge$6000) soft X-ray spectroscopy is feasible
within the scope of a modest mission. This resolving power would
open up new realms of physics to astronomy.

A rapid response mission to obtain high resolution soft X-ray and
Far-UV spectra of gamma-ray burst afterglows is also
possible. Such a mission would go well beyond current and planned
missions in determining the formation and nature of the warm-hot
intergalactic medium, the Cosmic Web.

\section{references}

\hspace{4mm} Aldcroft T.L., Elvis M., McDowell J.C., and Fiore
F., 1994, ApJ, 437, 584

Bechtold, J., Siemiginowska, A., Aldcroft, T., Elvis, M.
\& Dobrzycki, A. 2001, ApJ, 562, 133

Bloom, J.S., Kulkarni, S.R. \& Djorgovski, S.G. 2001, AJ, in
press, astro-ph/0010176

Cash, W. 1991, Applied Optics, 30-13, 1749

Castro, S., et al. 2001 ApJ, submitted, astro-ph/0110566

Cen, R., Ostriker, J.P. 1999a, ApJ, 514, 1


Dav\`e, R. et al. 2001, ApJ, 552, 473

Djorgovski, S.G. et al. 2001, 562, 654

Elvis, M. 2001, proceedings of `New Century of X-ray Astronomy',
Yokohama, Japan, astro-ph/0106053


Fiore, F., Nicastro, F., Savaglio, S., Stella, L. \& Vietri,
M. 2000 ApJL, 544, L7

Fiore, F.  2001, proceedings of `New Century of X-ray Astronomy',
Yokohama, Japan, astro-ph/0107276

In't Zand, J.J.M. et al. 2001, ApJ, 545, 266

Jha, S. et al. 2001, ApJL, 554, L155

Hellsten, U., Gnedin, N.Y., Miralda-Escud\`e, J. 1998, ApJ, 509,
56



Madau P., Pozzetti L. \& Dickinson M., 1998, ApJ, 498, 106

Masetti, N. et al., 2001, A\&A, 374, 382, astro-ph/0103296

Nicastro, F. et al. 2002a, ApJ, in press, astro-ph/0201058

Nicastro, F. et al. 2002b, {\em Nature}, submitted,
{\tt astro-ph/0208012}

Pareschi G. et al. 2001, NCXA Conf. 526.

Perna, R, \& Loeb, A. 1998, ApJL, 503, L135

Pettini, M., Smith, L.J., King, D.L. \& Hunstead, W. 1997, ApJ, 
486, 665

Pettini, M., Ellison, S.L., Steidel, C.C. \& Bowen, D.V. 1999,
ApJ, 510, 576

Prochaska, J.X. \& Wolfe, A.M. 1999, ApJS, 121, 369

Reeves J. et al. 2002, {\em Nature}, 416, 512
 
Savaglio, S., Fall, M., \& Fiore, F. 2002, ApJ, submitted


Sembach K. et al. 2002, ApJS, submitted, {\tt astro-ph/0207562}

Zappacosta L. et al. 2002, A\&A, in press {\tt astro-ph/0208033}
 
\end{document}